\documentclass[12pt, a4paper]{article}
\usepackage{amsfonts, amssymb}
\usepackage{graphicx}
\usepackage{latexsym,amsmath,color,array}

\usepackage{natbib}
\usepackage{enumerate}
\usepackage{bbding}
\usepackage{amssymb}
\usepackage{amsmath}
\usepackage{graphicx}
\usepackage{amsmath}
\usepackage{bbm}
\usepackage{mathtools}
 \usepackage{color}
 \usepackage{array}
 \usepackage{subfigure}
 \usepackage{multirow}
 \usepackage[dvipsnames]{xcolor}
 \usepackage{makecell}
 \usepackage{colortbl}

\def\T{{ \mathrm{\scriptscriptstyle T} }}
\newcommand{\m}{\phantom{-}}

\def\1{\mathbb{I}}

\newcounter{thm}[section]
\newcounter{appen}[section]
\newcounter{assum}[section]

\begin{document}

\title{Multi-Parameter Regression Survival Modelling:\\An Alternative to Proportional Hazards}
\author{Kevin Burke\footnote{Department of Mathematics and Statistics, University of Limerick, Limerick, Ireland; kevin.burke@ul.ie} \hspace{3cm}
Gilbert MacKenzie\footnote{CREST, Ensai, Rennes, France; gilbert.mackenzie@ul.ie} }
\date{}

\maketitle

\begin{abstract}
It is standard practice for covariates to enter a parametric model through a single distributional parameter of interest, for example, the scale parameter in many standard survival models. Indeed, the well-known proportional hazards model is of this kind. In this paper we discuss {a more general approach whereby covariates enter the model through \emph{more than one} distributional parameter simultaneously (e.g., scale \emph{and} shape parameters). We refer to this practice as ``multi-parameter regression'' {(MPR)} modelling and explore its use in a survival analysis context. We find that multi-parameter regression leads to more flexible models which can offer greater insight into {the underlying data generating process}. To illustrate the concept, we consider the two-parameter Weibull model which leads to time-dependent hazard ratios, thus relaxing the typical proportional hazards assumption and motivating a new test of proportionality. A novel variable selection strategy is introduced for such multi-parameter regression models. It accounts for the correlation arising between the estimated regression coefficients in two or more linear predictors -- a feature which has not been considered by other authors in similar settings. The methods discussed have been implemented in the \texttt{mpr} package in \texttt{R}.}

\smallskip


\end{abstract}

\qquad

\newpage

\section{Introduction}
The standard approach to (parametric) regression is to relate covariates to one parameter of specific interest, for example, consider generalized linear models \citep{mcculneld:1989} where covariates enter through the location parameter while the dispersion parameter is simply a constant. We will refer to this standard approach as \emph{single parameter regression} (SPR). The question we ask is why any \emph{one} parameter should be deemed the ``interest'' parameter in terms of covariate analysis? This standard SPR approach neglects the potential influence of covariates on other distributional parameters which may also be important in describing the phenomenon under study. A more flexible approach is to allow covariates to enter the model through \emph{multiple} distributional parameters, e.g., location \emph{and} dispersion simultaneously. Hereafter, we refer to this approach as \emph{multi-parameter regression} (MPR).

Multi-parameter regression modelling has appeared several times in the literature. In the context of normal linear regression, \citet{park:1966} and \citet{harvey:1976} modelled the dispersion parameter as a function of covariates to address heteroscedasticity, while \citet{smyth:1989} modelled dispersion in the more general case of generalized linear models. More recently, structured dispersion models have been considered by \citet{leeneld:2001, leeneld:2006}. The generalized additive models for location, scale and shape (GAMLSS) framework \citep{rigbystas:2005, stasrigby:2007} goes beyond the exponential family to include a large variety of distributions in which covariates may enter through multiple parameters simultaneously; see also http://www.gamlss.org. The MPR approach has also been considered in longitudinal data analysis (joint mean-covariance models) \citep{panmac:2003,panmac:2006,panmac:2007} and in quantile regression \citep{noufjones:2013}.

In the setting of survival analysis, fully parametric regression models have been much less popular than Cox's \citeyearpar{cox:1972} semi-parametric proportional hazards (PH) model where the hazard function is given by $\lambda(t\,|\,x) = \exp(x^\T \beta) \lambda_0(t)$
where $x = (x_{1}, \ldots, x_{p})^\T$ and $\beta = (\beta_1, \ldots, \beta_p)^\T$ are vectors of covariates and regression coefficients, respectively, and $\lambda_0(t)$ is a non-parametric function. 
Due to its wide acceptance of and availability in standard software, the Cox model is often fitted to data where the proportional hazards assumption does not hold \citep{schemp:1992}. 
Although the Cox model can be generalized in various ways 
\citep{mcgilchristaisbett:1991, gray:1992, klein:1992, nielsenetal:1992, gramther:1994, therngram:2000, martetal:2002, tianetal:2005}, our focus will be on fully parametric multi-parameter regression models.

We note that the basic Cox model is in fact a single parameter regression model since the \emph{scale} of $\lambda(t\,|\,x)$ is modelled via the component $\exp(x^\T \beta)$ whereas its \emph{shape} has no specific structure, being absorbed entirely by $\lambda_0(t)$. Although the stratified Cox model is somewhat analogous to a multi-parameter regression model in the sense that covariates enter the shape of $\lambda_0(t)$, this is limited to categorical covariates.
Moreover, since $\lambda_0(t)$ is not modelled directly within the Cox model,  we cannot estimate the effects of such stratification variables on the hazard. 
However, if instead we consider a \emph{fully} parametric hazard, with parameters controlling its scale and shape, it is straightforward to develop a model where covariates enter through these scale \emph{and} shape parameters. Indeed, many popular survival distributions have two such parameters, for example, Weibull, log-normal, gamma and log-logistic \citep{kalbprent:2002,lawless:2003}.  Within these two-parameter survival distributions, covariates are typically introduced via the scale parameter leading to proportional hazards (PH), accelerated failure time (AFT) or proportional odds (PO) models depending on the distribution. However, further flexibility could clearly be achieved by modelling the scale \emph{and} shape parameters (i.e., \emph{multi}-parameter regression), thereby  accommodating a wider variety of survival data.

Multi-parameter regression is not wholly novel in a survival setting; early references include \citet{taulb:1979} and \citet{gay:1987} who used polynomial and Gompertz hazard models respectively. \citet{anderson:1991} considered a log-linear model for the survival time where both the location and dispersion parameters depended on covariates (generalizing the AFT model) and, as an example, applied a Weibull MPR model (albeit different from the one in this paper) to the Framingham Heart Study data;  interestingly, this model has been adopted in coronary heart disease literature where it is known as a ``Framingham equation'' \citep{mcewanetal:2004}.
More recently, the MPR approach has been used in the context of the inverse Gaussian model \citep{leewhit:2006,aalen:2008,leewhit:2010}. Furthermore, in their development of semi-parametric maximum likelihood estimation, \citet{zenglin:2007} considered heteroscedastic transformation models for survival data encompassing multi-parameter regression. It is also noteworthy that cure models \citep{farewell:1982, mallerzhou:1995, lambertetal2007, decastroetlal:2010} fall within the MPR framework, i.e., covariates typically enter via the scale parameter and the cure probability.

Notwithstanding these developments, we note that multi-parameter regression in survival has not previously received much attention as a general procedure and is certainly not in mainstream use. We therefore explore and discuss the consequences of the multi-parameter regression approach using the popular Weibull model and show its flexibility in real data analysis. In particular,  the MPR model provides a useful alternative to the proportional hazards assumption and produces a new test of deviation from proportionality. It may be fitted using our new \texttt{mpr} package \citep{burke:2016} in \texttt{R} which implements the techniques developed in this paper. See also the \texttt{gamlss.cens} package \citep{stasrigby:2016} which extends the GAMLSS framework to survival data.

The paper is organized as follows. In \S \ref{mprmods} we introduce and develop the Weibull MPR model. Hypothesis testing and variable selection are discussed, including a simulation study to assess our proposed MPR selection procedure, in \S \ref{hyptestsec}. An application of the methodology to lung cancer data 
appears in \S \ref{datasec} and, finally, we conclude with some remarks in \S \ref{discuss}.

\section{The Weibull multi-parameter regression model \label{mprmods}}
The Weibull distribution is one of the most popular survival distributions and its hazard function is given by
\begin{align}
\lambda(t) &= \lambda \gamma t^{\gamma - 1}
\end{align}
where $\lambda > 0$, the scale parameter, controls the overall magnitude of the hazard and $\gamma > 0$, the shape parameter, controls its time-evolution; it can increase ($\gamma > 1$), decrease ($\gamma < 1$) or remain constant ($\gamma = 1$) over time. The Weibull MPR model is generated by introducing covariates into \emph{both} parameters as follows:
\begin{align}
\log(\lambda) &= x^\T \beta, \quad &\log(\gamma)  &=  z^\T \alpha, \label{mprsweib}
\end{align}
where the log-link is used to ensure positivity of both parameters, $x = (1, x_{1}, \ldots, x_{p})^\T$ and $z = (1, z_{1}, \ldots, z_{q})^\T$ are scale and shape covariate vectors which may or may not contain covariates in common and $\beta = (\beta_0, \beta_1, \ldots, \beta_p)^\T$ and $\alpha = (\alpha_0, \alpha_1, \ldots, \alpha_q)^\T$ are the corresponding regression coefficients. Maximum likelihood estimation of the regression coefficients is straightforward and has been implemented in the \texttt{mpr} package \citep{burke:2016}.

Under the Weibull MPR model, the hazard ratio for a binary covariate, $c$, common to both scale and shape regression components, is
\begin{equation}
\frac{\lambda(t\,|\,c = 1)}{\lambda(t\,|\,c = 0)}
= \exp(\beta_c + \alpha_c) \,\,t^{\exp(\tilde{z}^\T \alpha)\{\exp(\alpha_c) - 1\}}, \label{weibhr}
\end{equation}
where $\beta_c$ and $\alpha_c$ are the scale and shape coefficients of $c$ respectively and $\tilde{z}^\T \alpha = z^\T \alpha - c\,\alpha_c$ represents all of the other terms in the shape linear predictor excluding $c$. The hazard ratio is time-dependent where the time-evolution depends on the other shape-covariates via $\tilde{z}$. This dependence on $\tilde{z}$ can be dealt with by inserting a ``typical'' shape covariate profile 
or by integrating (\ref{weibhr}) over the empirical distribution of $\tilde{z}$ in the dataset (cf. \citet{neldlane:1982, kar:1987, shenflem:1997, martpipper:2013}). Most importantly, it is the sign of $\alpha_c$ which solely determines the path of the hazard ratio. 
When $\alpha_c=0$, the ratio of hazards reduces to the usual PH constant, $\exp(\beta_c)$. Thus, the Weibull MPR model directly extends the proportional hazards model, providing a new test of proportionality adjusted for other scale and shape covariates. Furthermore, for $\alpha_c \ne 0$, setting (\ref{weibhr}) equal to one and solving yields an explicit solution for the time-point at which crossing hazards occur: $t_c = \exp\left\{(\beta_c + \alpha_c) [\exp(\tilde{z}^\T \alpha)\{1-\exp(\alpha_c)\}]^{-1} \right\}$.

We note that \citet{anderson:1991}  considered a different Weibull MPR model. \citeauthor{anderson:1991}'s model is based on an alternative hazard parametrization, $\lambda(t) = \lambda \gamma (\lambda t)^{\gamma - 1}$, more suited to AFT modelling. One easily finds that it implies a hazard ratio whose functional form is somewhat more complicated than (\ref{weibhr}) and which does not reduce to $\exp(\beta_c)$ when $\alpha_c = 0$. Thus, the model presented here provides a more direct extension to the PH model. However, just as our model produces a PH test for a given covariate, \citeauthor{anderson:1991}'s model produces an analogous AFT test -- but this was not explored in \citet{anderson:1991}. Furthermore, the primary model emphasized by \citet{anderson:1991} is a restricted and asymmetric parametrization in which the shape component is related to the scale component via  $\log \gamma = \theta_0 + \theta_1 \log \lambda$ (although the unrestricted parametrization was also developed and applied further in \citet{odell:1994}). Such parametrizations do not permit variable selection in both regression components, a  valuable procedure  which enables one to determine which covariates have non-PH effects. The identification of such variables is important in practice (see Section \ref{hyptestsec}). Finally, it is worth highlighting that, although both we and \citet{anderson:1991} consider the Weibull distribution as an example, the MPR approach is not limited to any one distribution and, indeed, the accompanying \texttt{mpr} package \citep{burke:2016} includes a variety of popular survival distributions.

\section{Hypothesis testing and variable selection\label{hyptestsec}}

We now consider the implications of correlation between estimated regression coefficients within MPR models on hypothesis testing and variable selection strategies. These issues do not appear to have been considered in depth by other authors.

\subsection{Hypothesis testing}
In standard \emph{single}-parameter regression (SPR) survival models, covariates only appear in the scale and, therefore, the importance of a particular covariate, $c$, is determined by testing if its regression coefficient, $\beta_c$, differs significantly from zero. However, in the MPR model there are two hypotheses of interest, namely: (i) $H_0: \beta_c = 0$ and (ii) $H_0: \alpha_c = 0$. Furthermore, the estimated coefficients for the same covariate in different regression components (i.e.,  $\hat{\beta}_c$ and $\hat{\alpha}_c$) are typically quite correlated (see Appendix A). The consequence of this higher correlation is that the scale effect,  $\hat{\beta}_c$,  is most heavily adjusted for the shape effect, $\hat{\alpha}_c$, and vice versa. Thus, we recommend that covariates should be considered in both regression components simultaneously, since we can imagine scenarios where a covariate only becomes significant when it is present in both components.

{It is also of interest to determine the total effect of $c$,} i.e., by testing (iii) $H_0: \beta_c = \alpha_c = 0$. Using the asymptotic result $(\hat{\beta}_c,\hat{\alpha}_c)^T \sim N[(\beta_c,\alpha_c)^T, \Sigma_{\hat{\beta}_c,\hat{\alpha}_c} ]$, where $\Sigma_{\hat{\beta}_c,\hat{\alpha}_c}$ is the relevant $2 \times 2$ covariance sub-matrix of $\Sigma(\hat{\theta})$, we have that
\begin{equation}
[(\hat{\beta}_c,\hat{\alpha}_c)^T - (\beta_c,\alpha_c)^T ]^T \,\, \Sigma^{-1}_{\hat{\beta}_c,\hat{\alpha}_c} \,\, [ (\hat{\beta}_c,\hat{\alpha}_c)^T - (\beta_c,\alpha_c)^T ] \, \sim \, \chi^2_2.  \label{confell}
\end{equation}
{Thus, a p-value can be obtained by setting $(\beta_c,\alpha_c)^T = (0, 0)^T$ in (\ref{confell}) and comparing this statistic to the $\chi^2_2$ distribution. Furthermore, a $(1-\alpha) 100 \%$ confidence ellipse for $(\beta_c,\alpha_c)^T$ is given by the set of $(\beta_c,\alpha_c)^T$ points defining the contour line such that (\ref{confell}) is equal to $\chi^2_{2,\,1-\alpha}$ (see \citet{friendlyetal:2013} for a detailed account of confidence ellipses).}

\subsection{Selection procedure}

We now define $M(x,z)$ {to be} a model with scale and shape covariate vectors $x$ and $z$ respectively. Hence, let $M_0 = M(\tilde{x},\tilde{z})$ be the model where $c$ does not appear (i.e., $c\not\in \tilde{x}, \tilde{z}$). Now, there are three models which include $c$: $M_{{\beta_c}}=M(\tilde{x} \, \cup \, c,\tilde{z})$, $M_{{\alpha_c}}=M(\tilde{x},\tilde{z}\,\cup\,c)$ and $M_{{\beta_c\alpha_c}}=M(\tilde{x}\,\cup\,c,\tilde{z}\,\cup\,c)$, {respectively}. Therefore, we can test hypotheses (i), (ii) and (iii) above by comparing $M_{\beta_c\alpha_c}$ with $M_{\alpha_c}$, $M_{\beta_c\alpha_c}$ with $M_{\beta_c}$ and $M_{\beta_c\alpha_c}$ with $M_0$, respectively. {This can be done} by means of likelihood ratio tests or information criteria (see \citet{burnand:2002}); this approach is more in line with variable selection procedures {\citep{miller:2002}}. Of course inferential problems associated with variable selection methods are well documented \citep{miller:1984, hurvtsai:1990, zhang:1992} and automatic selection procedures are much criticized. Nonetheless, such procedures are useful when the number of covariates is large, so that methods for {MPR} models are required. Accordingly, an algorithm for MPR forward selection is described in Appendix B which also includes a comparison with the \texttt{gamlss} stepwise procedure.

\subsection{Simulation study\label{simsec}}

We now evaluate the performance of {a stagewise} MPR variable selection {procedure which involves starting from the null model, contains both forward and backward (scale, shape and simultaneous) steps and uses information criteria (AIC and BIC) as the basis of selection.

We simulated data from the Weibull MPR model with
\begin{align*}
\log \lambda &= x^T \, (-1.5,-1.0,\m1.0,\m0.5,-0.5,\m0.0,\m0.0,-0.8,\m0.5,\m0.0,\m0.0)^T, \\[0.3cm]
\log \gamma &=  z^T \, (\m0.5,\m0.4,-0.4,\m0.2,-0.2,\m0.4,-0.2,\m0.0,\m0.0,\m0.0,\m0.0)^T,
\end{align*}
where $x = z = (1, x_1, \ldots, x_{10})^T$ is a vector of independent binary variables. Thus, the first four covariates, $x_1$, $x_2$, $x_3$ and $x_4$, affect both the scale and the shape, $x_5$ and $x_6$ affect the shape only, $x_7$ and $x_8$ affect the scale only and, finally, $x_9$ and $x_{10}$ have no affect on either component. This setup gives a good variety of scale and shape effects of different strengths where the coefficient values were chosen to lead to realistic survival data. Furthermore, we varied the sample size ($n = $ 100, 500 and 1000) and censored proportion ($p = $ 20\%, 50\% and 80\%) giving 9 scenarios in total, each of which was repeated 500 times.

{At the $j$th repetition of a particular scenario, $j=1,\ldots,500$, the selection procedure was applied: let $x^{(j)}$ and $z^{(j)}$ denote the sets of scale and shape covariates selected in this repetition. Over these 500 simulation replicates, the relative (scale and shape) selection frequencies for the covariate $x_k$, $k=1,\ldots,10$, are given by}
\begin{align*}
\hat{\pi}_{x_k \in\, x} &= \frac{1}{500}\sum_{j=1}^{500} 1(x_k \in x^{(j)}) \text{\hspace{0.8cm}and\hspace{-0.8cm}} & \hat{\pi}_{x_k \in\, z} &= \frac{1}{500}\sum_{j=1}^{500} 1(x_k \in z^{(j)})
\end{align*}
{respectively. The relative frequencies, for selection based on AIC and BIC, respectively, in each of the 9 simulation scenarios, are displayed in Fig.~\ref{freqselectfig}.}

\begin{figure}[!htb]
\centering
\begin{tabular}{c@{\hspace{0.1cm}}c}
\includegraphics[width=0.5\textwidth, trim = 0.3cm 2.59cm 1cm 1cm, clip]{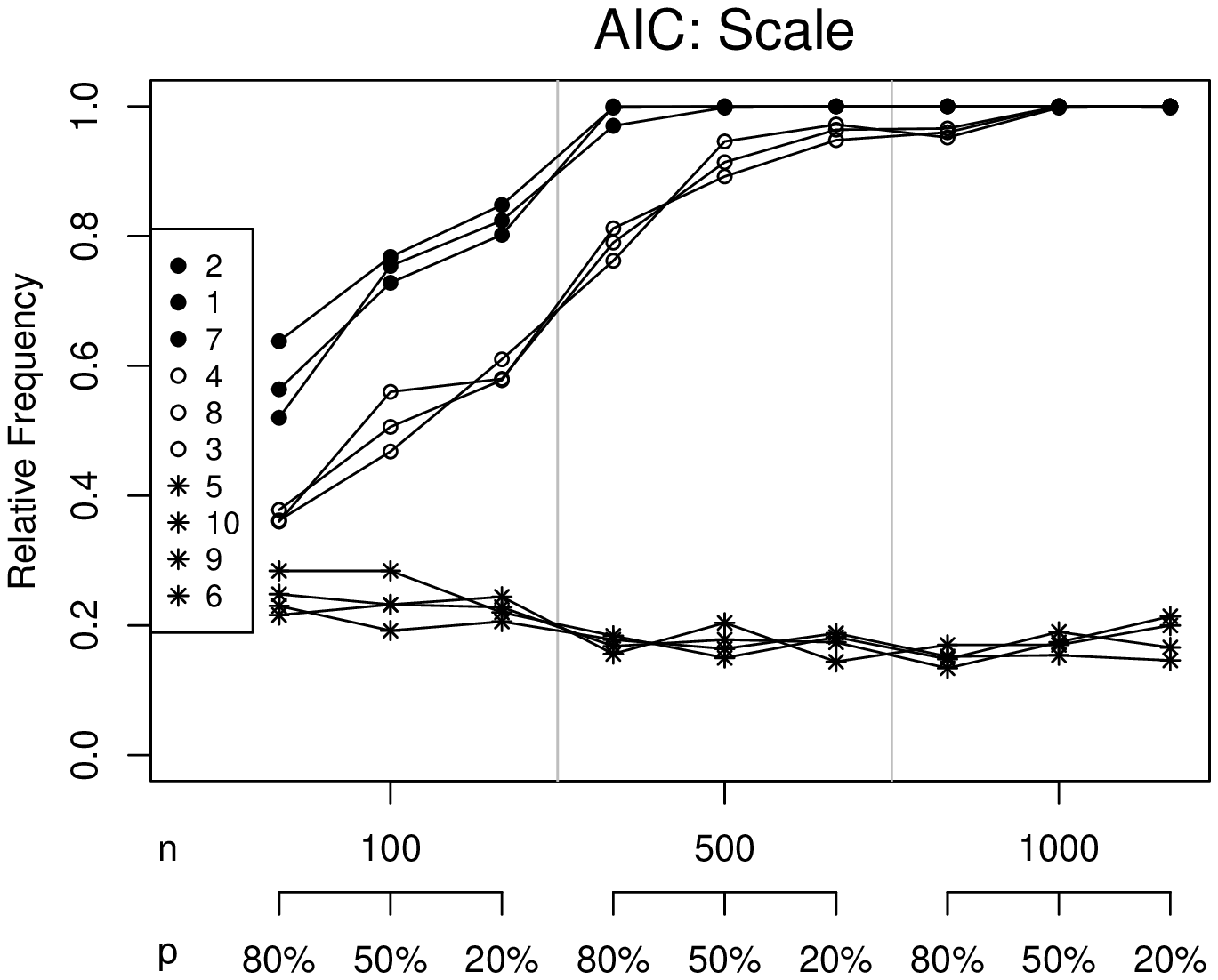} &
\vspace{0.6pt}\includegraphics[width=0.437\textwidth, trim = 2.05cm 2.59cm 1cm 1cm, clip]{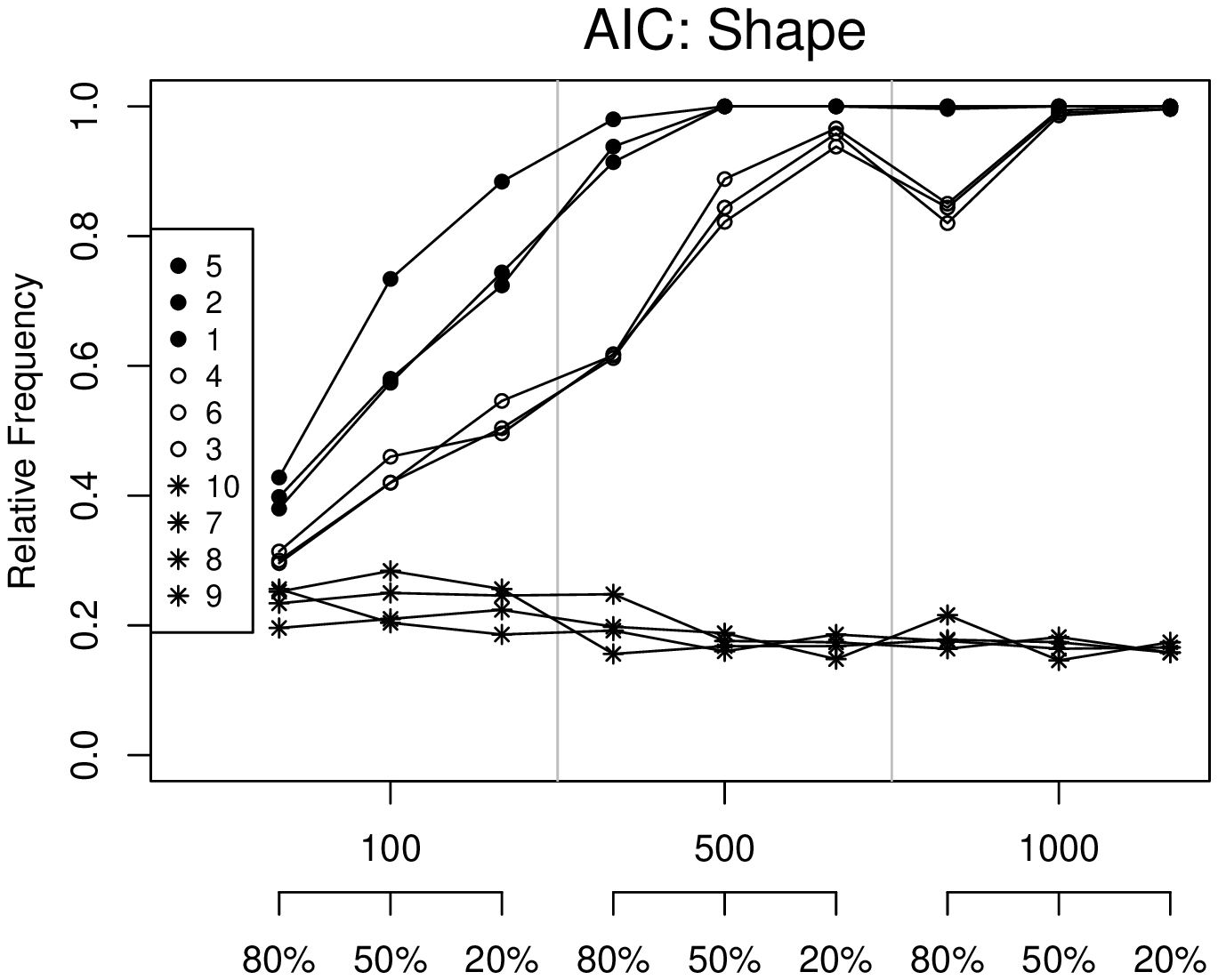}\\[-0.1cm]
\includegraphics[width=0.5\textwidth, trim = 0.3cm 0.4cm 1cm 1cm, clip]{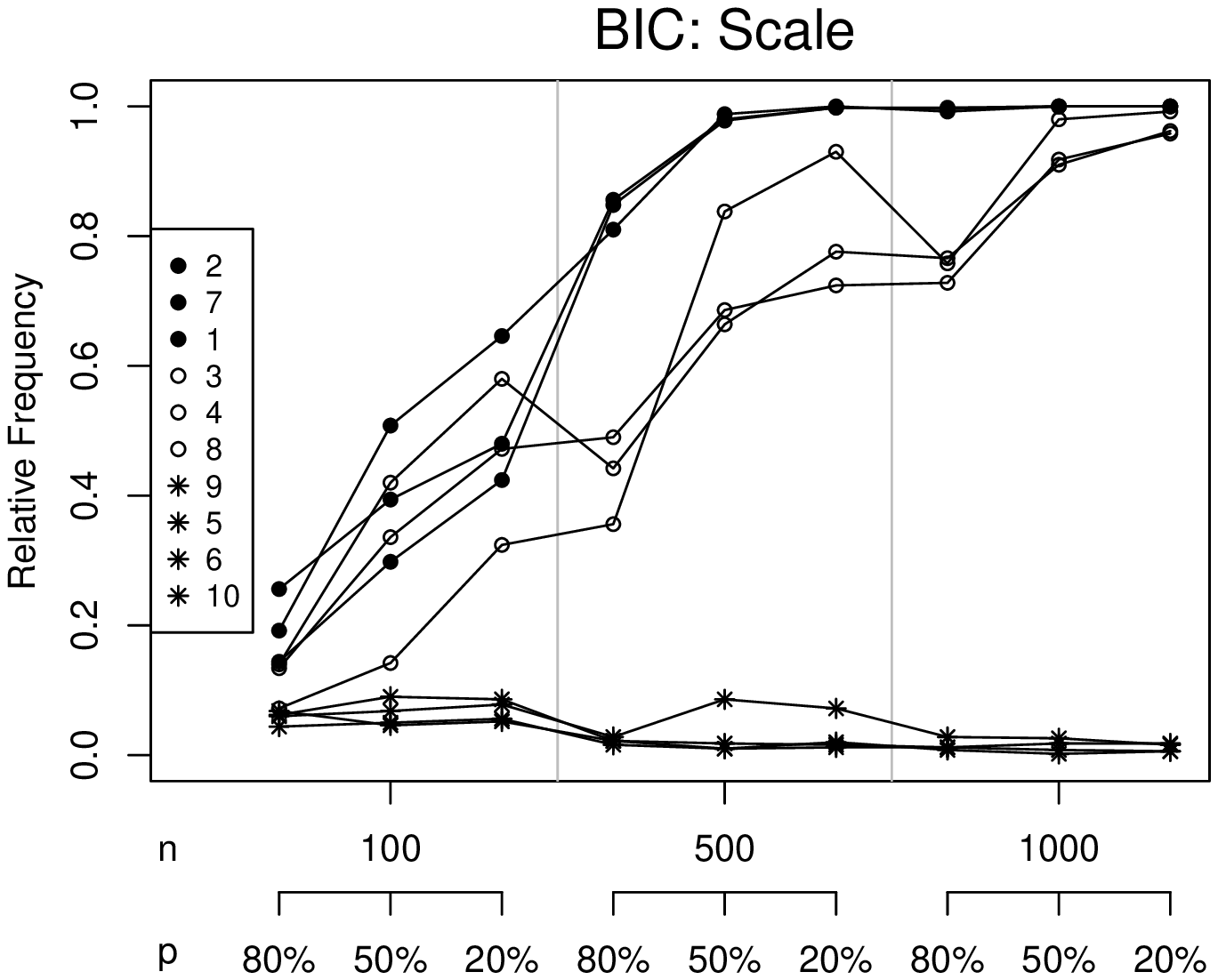} &
\vspace{0.6pt}\includegraphics[width=0.437\textwidth, trim = 2.05cm 0.4cm 1cm 1cm, clip]{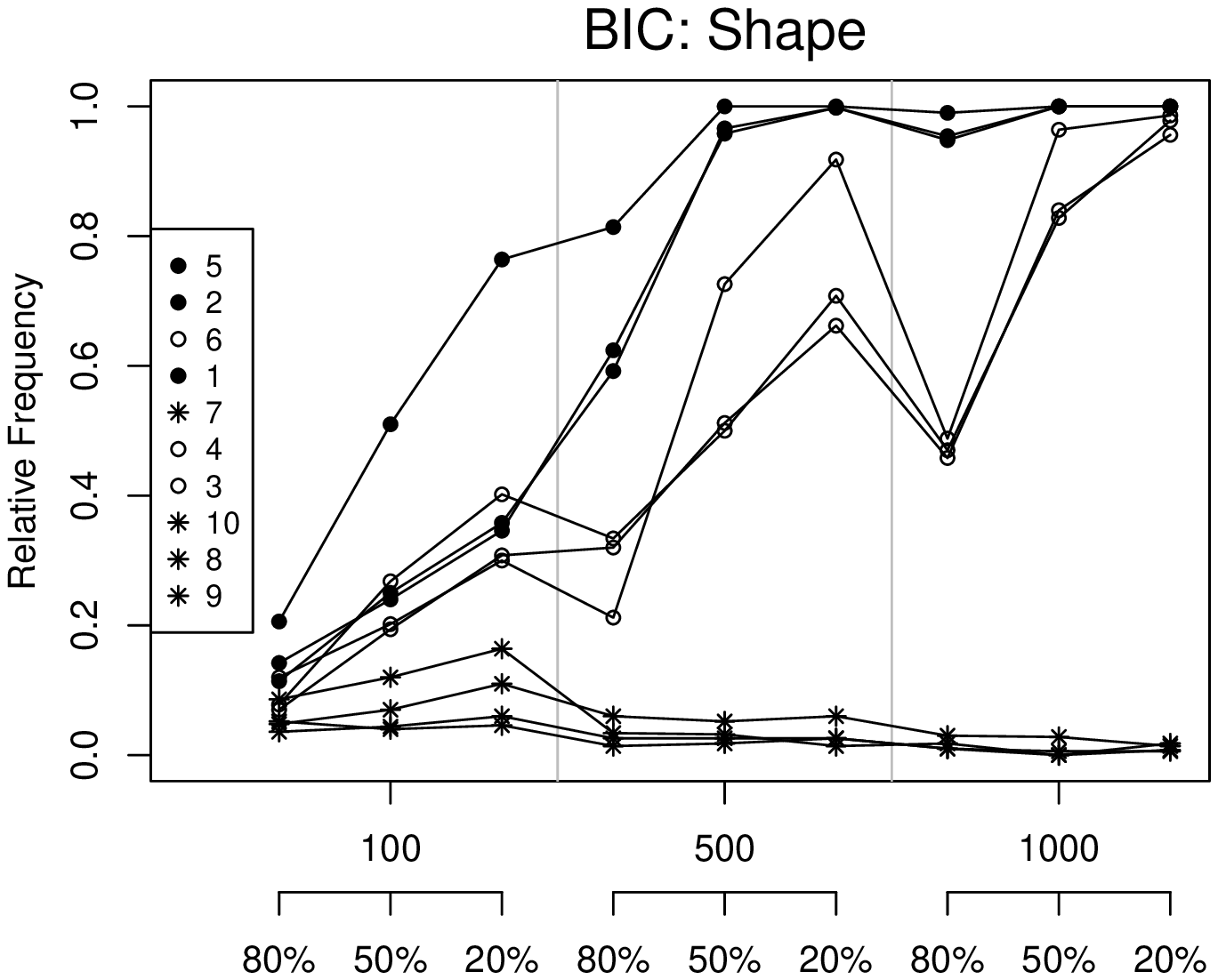}\\[0.6cm]
\end{tabular}
\caption{Selection procedure based on AIC (top) and BIC (bottom). Relative selection frequency is shown for the scale (left) and shape (right) components over each simulation scenario for each covariate (stronger effect = solid circle, weaker effect = open circles, no effect = asterisk). The order of terms in the legend matches the order from scenario one, $(n,p) = (100, 80\%)$, e.g., using AIC for the scale, $x_2$ and $x_6$ have the highest and lowest selection frequencies, respectively, in scenario one. \label{freqselectfig}}
\end{figure}

We first consider the results for selection based on AIC (Fig.~\ref{freqselectfig}, top panel). It is clearly difficult to determine which covariates affect survival when the information available is low, i.e., $(n,p)  = (100, 80\%)$. However, even when the sample size is small, the covariates with stronger effects (i.e., those with larger regression coefficients) are still selected 75\% - 85\% of the time provided that censoring is low ($p = 20\%$). When a moderate level of information is available, $(n, p) = (500, 50\%)$, the covariates with stronger effects are selected 100\% of the time while those with weaker effects are selected over 80\% of the time. Furthermore, as the sample size increases, the relative frequency of selection of any covariate with nonzero effect approaches one. Although important covariates were identified quite successfully, covariates with no effect were selected approximately 20\% of the time across all scenarios using AIC. This can easily be combatted by using BIC (Fig.~\ref{freqselectfig}, bottom panel) but it comes at the expense of identifying important covariates less often than when using AIC (particularly in low information scenarios) which is not an unexpected finding.

\section{Analysis of lung cancer data \label{datasec}}
Here we analyse a lung cancer dataset which appeared in a 1995 Queen's University Belfast PhD thesis by P.~Wilkinson and was further analysed by \citet{mackenzie:1996}.  The data contains individuals diagnosed with lung cancer in Northern Ireland during the period October 1st 1991 to September 30th 1992. Time was measured, in months, from date of diagnosis until the earlier of the occurrence of death or the study end date, which was May 30th 1993. In this observational study there were 855 individuals, of whom 673 died and the remainder were right-censored.  A number of (categorical) covariates were measured on each individual, namely: treatment type, age group, WHO status, sex, smoking status, cancer cell type, metastases, and albumen and sodium levels in the blood.

We analyse the data using the Weibull MPR model from the \texttt{mpr} package. Single factor and multi-factor analyses are presented to illustrate both the capabilities of the MPR approach 
and the application of MPR variable selection. It is worth noting that a number of the categorical variables have missing values; in these cases, an extra category labelled ``Missing'' was created. Although this ad-hoc approach retains observations in the analysis, it can lead to bias in the estimated parameters \citep{hortanandkleinman:2007,jones:1996,vachblettner:1991}. However, addressing missingness is not our aim here.

\subsection{Single factor analysis: treatment \label{treatsec}}

{We first investigate the unadjusted} effect of treatment which has five levels, namely: palliative care (the reference category), surgery, chemotherapy, radiotherapy and chemotherapy + radiotherapy combined. {Two models were fitted to the data: the \emph{multi-parameter regression} Weibull model, and the standard, but more restrictive, shape-constant \emph{single parameter regression} model, i.e., the \emph{proportional hazards} Weibull model. These models (labelled MPR and PH) are summarized in Table \ref{restable}.}

\begin{table}[htb]
\caption{Single factor analysis results: standard errors are in brackets. \label{restable}}
\centering
\begin{tabular}{l@{\hspace{0.8cm}}r@{\hspace{0.2cm}}c@{\hspace{0.5cm}}r@{\hspace{0.2cm}}c@{\hspace{0.8cm}}r@{\hspace{0.2cm}}c@{\hspace{0.5cm}}r@{\hspace{0.2cm}}c}
\hline
&&&&&&&&\\[-0.4cm]
               &   \multicolumn{4}{c}{\hspace{-1.0cm}MPR\hspace{-0.7cm}}  & \multicolumn{4}{c}{\hspace{-0.2cm}PH\hspace{-0.7cm}}  \\
\hline
&&&&&&&&\\[-0.3cm]
               &  \multicolumn{2}{c}{\hspace{-0.5cm}Scale\hspace{-0.2cm}} & \multicolumn{2}{c}{\hspace{-0.8cm}Shape\hspace{-0.2cm}} & \multicolumn{2}{c}{\hspace{-0.5cm}Scale\hspace{-0.2cm}} & \multicolumn{2}{c}{\hspace{0cm}Shape\hspace{-0.2cm}} \\
Intercept      &  -1.28 & (0.08)  &  -0.19 & (0.04)   &  -1.48  & (0.08)  &  -0.07  & (0.03)    \\
Palliative     &   0.00 & ------  &   0.00 & ------   &   0.00  & ------  &  ------ &  ------    \\
Surgery        &  -3.91 & (0.83)  &   0.59 & (0.20)   &  -2.22  & (0.23)  &  ------ &  ------   \\
Chemo          &  -0.50 & (0.32)  &   0.07 & (0.14)   &  -0.40  & (0.17)  &  ------ &  ------   \\
Radio          &  -1.26 & (0.19)  &   0.34 & (0.07)   &  -0.57  & (0.09)  &  ------ &  ------   \\
C+R            &  -4.06 & (0.88)  &   0.97 & (0.16)   &  -0.87  & (0.20)  &  ------ &  ------   \\
&&&&&&&&\\[-0.3cm]
\hline
&&&&&&&&\\[-0.3cm]
$\ell(\hat{\theta})$ & \multicolumn{4}{c}{\hspace{-1.0cm}-1938.1\hspace{-0.7cm}}  & \multicolumn{4}{c}{\hspace{-0.2cm}-1960.8\hspace{-0.7cm}}    \\
AIC & \multicolumn{4}{c}{\hspace{-1.0cm}\m3896.2\hspace{-0.7cm}}  & \multicolumn{4}{c}{\hspace{-0.2cm}\m3933.5\hspace{-0.7cm}}    \\
BIC & \multicolumn{4}{c}{\hspace{-1.0cm}\m3943.8\hspace{-0.7cm}}  & \multicolumn{4}{c}{\hspace{-0.2cm}\m3962.0\hspace{-0.7cm}}    \\
\hline
\end{tabular}
\end{table}

In the PH model, the $\hat \beta$ coefficients are negative which means that treatment reduces the hazard (compared to palliative care). All treatments are statistically significant (at the 5\% level) and, furthermore, the relative merit of each treatment is easily determined from the magnitude of the coefficients. However, the situation is more complex in the MPR model. The $\hat \beta$ coefficients are also negative in this case but the $\hat \alpha$ coefficients are positive. Thus, although the hazard is reduced through treatment, it is increasing relative to palliative care over time. In other words, the effectiveness of each treatment reduces over time.  All treatments are statistically significant in both the scale and shape components apart from chemotherapy (neither component). This latter statement is based on considering the scale and shape effects separately; joint effects can be assessed using (\ref{confell}). To this end, joint confidence ellipses for the regression coefficients are shown in Fig.~\ref{ellipse95} along with the individual confidence intervals for comparison. Although the chemotherapy effect is non-significant, its confidence ellipse only just covers the $(0,0)$. Furthermore, while zero is well within the $\hat\alpha_{\text{C}}$ confidence interval, the $\hat\beta_{\text{C}}$ confidence interval only just includes zero. This motivates consideration of a reduced MPR model, fixing $\alpha_{\text{C}}=0$. Indeed, we find that this reduced MPR model (not shown) has improved AIC (3894.5) and BIC (3937.2), respectively and, interestingly, we then get $\hat\beta_{\text{C}}=-0.37~ (\text{S.E. }= 0.17)$, i.e., very close to the PH analysis, while all other coefficients are virtually unchanged from the original MPR model. This latter finding highlights that the MPR analyis permits the inclusion of simpler PH effects (as expected from Section \ref{mprmods}), however, for the sake of brevity, we will not consider the reduced model further.

\begin{figure}[!htb]
\centering
\includegraphics[width=1\textwidth, trim = 0.0cm 0.2cm 0.5cm 0.5cm, clip]{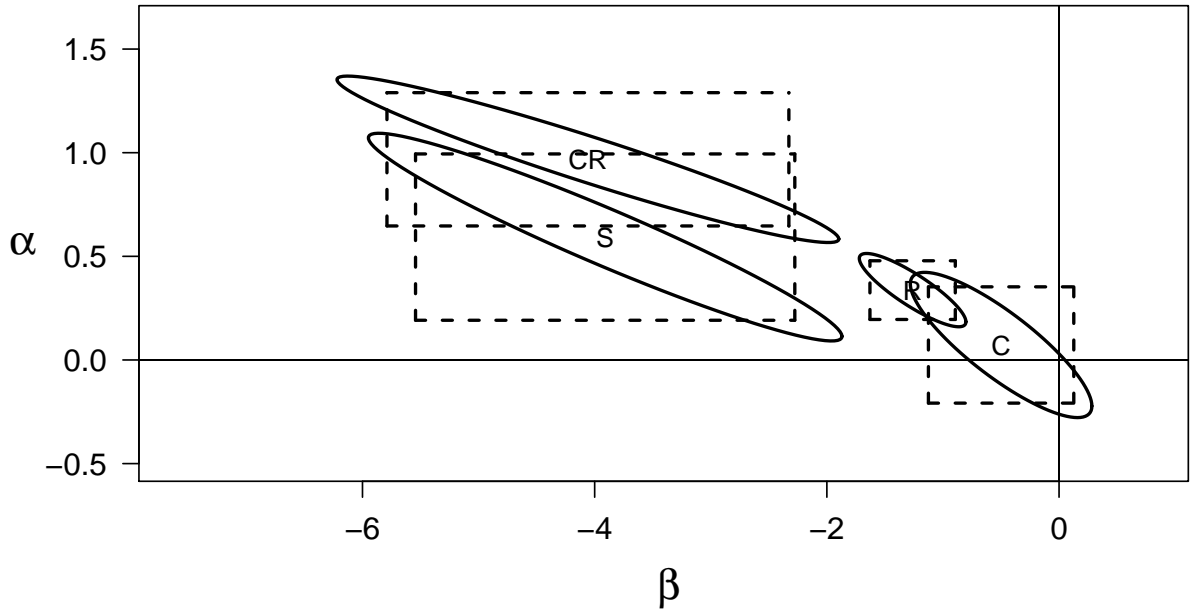}
\caption{Single factor analysis: joint 95\% confidence ellipses (solid) and individual 95\% confidence intervals (dash) for the pairs $(\beta_C, \alpha_C)$, $(\beta_R, \alpha_R)$, $(\beta_{CR}, \alpha_{CR})$ and $(\beta_S, \alpha_S)$ where C = chemotherapy, R = radiotherapy, CR = chemotherapy + radiotherapy and \mbox{S = surgery} respectively.\label{ellipse95}}
\end{figure}

{Although the $\beta$- and $\alpha$-coefficients in the MPR model provide some preliminary insights, the hazard ratios, (\ref{weibhr}), combine information from both of these components which allows us to determine the overall treatment effects}. Thus, Fig.~\ref{hrplot} shows the hazard ratio for each treatment relative to palliative care along with along with 95\% confidence intervals calculated using the delta method. For comparison, the PH hazard ratios and confidence intervals are also shown. Based on the MPR analysis, the only treatment superior to palliative care over the full range of time is surgery. Chemotherapy appears to have a slight early effect in the first 4 months. Radiotherapy and the combined treatment are both superior to palliative care initially, but their effectiveness wears off after about 7 months. Although there appears to be some evidence that the combined treatment may become more hazardous than palliative care later in time, we note that there is much less information in the tails (15 - 20 months) and, indeed, this could also be a consequence of different frailties in the groups. Of course, being an observational study, these are not the true treatment effects since they depend on the net interplay of other factors, for example, many of the individuals selected for surgery were younger, healthier and free from metastases.
\begin{figure}[!htb]
\centering
\includegraphics[width=1\textwidth, trim = 0.2cm 0.2cm 0.5cm 0.5cm, clip]{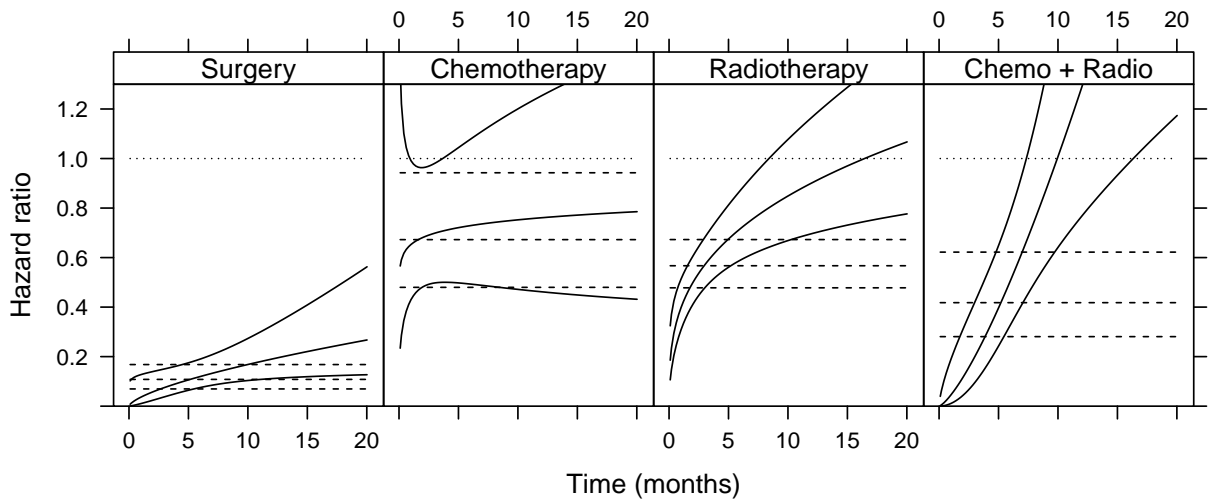}
\caption{Single factor analysis: treatment hazard ratios (reference = palliative care) with 95\% confidence intervals for MPR (solid) and PH (dash) models with line of equality (dot).\label{hrplot}}
\end{figure}

Figure \ref{kmfitplot} shows the Kaplan-Meier survivor curves \citep{kapmeier:1958} for each treatment group along with the model-based (MPR and PH) predicted survivor curves calculated using the well-known relationship $S(t) = \exp(-\int_0^t \lambda(u) du)$ where $S(t)$ and $\lambda(t)$  are the survivor and hazard functions respectively. We can see that the MPR model fits the data more closely than the PH model. More formally, {this improvement in fit is confirmed} by carrying out a likelihood ratio test $(p \ll 0.001)$ and {the fact that the MPR model has much lower} AIC and BIC values. It is noteworthy that the regression coefficients, standard errors and predicted survivor curves arising from {the semi-parametric} Cox model (not shown) are almost identical to {those of} the PH Weibull model.
\begin{figure}[!htb]
\centering
\includegraphics[width=1\textwidth, trim = 0.2cm 0.2cm 0.5cm 0.5cm, clip]{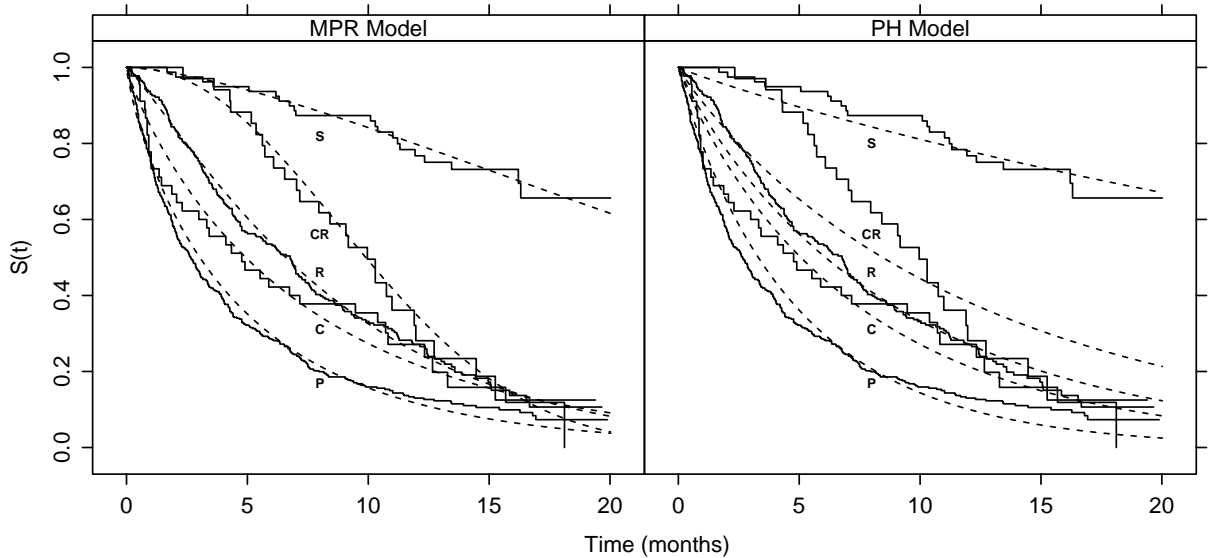}
\caption{Single factor analysis: Kaplan-Meier (step, solid) curves with model based curves (smooth, dash) overlayed where P = palliative, C = chemotherapy, R = radiotherapy, CR = chemotherapy + radiotherapy and S = surgery respectively.\label{kmfitplot}}
\end{figure}

\subsection{Multi-factor analysis: variable selection}

{We now consider a full covariate analysis of the lung cancer data. In order to initially determine the importance of each covariate, first the full model (scale and shape components saturated) was fitted to the data; this model had $\ell(\hat\theta) = -1798.44$ and AIC $=3700.88$ respectively. Then, 27 reduced models were fitted which arise through excluding each of the nine covariates from scale, the shape and simultaneously from the scale and shape. These models were compared to the full model using likelihood ratio tests and AIC differences.}

\begin{table}[!htbp]
\caption{Likelihood Ratio Tests and AIC differences \label{tableLRandAIC}}
\centering
\begin{tabular}{l@{\hspace{0.8cm}}rrr@{\hspace{0.8cm}}rrr}
\hline
&&&&&&\\[-0.4cm]
         &  \multicolumn{3}{c}{\hspace{-0.3cm}L.R. Test (p-value)} & \multicolumn{3}{c}{\hspace{-0.3cm}$\Delta_{\text{AIC}}$} \\
\hline
&&&&&&\\[-0.4cm]
 &  \multicolumn{1}{c}{Scale} & \multicolumn{1}{c}{Shape}  & \multicolumn{1}{c}{\hspace{-0.6cm}Joint} & \multicolumn{1}{r}{Scale} & \multicolumn{1}{r}{Shape}  & \multicolumn{1}{r}{Joint} \\
\hline
&&&&&&\\[-0.3cm]
Treatment         & $<\!0.001$ & $<\!0.001$ & $<\!0.001$ &  32.4 & 14.8 & 60.4 \\
Age Group         & 0.227     & 0.039     & 0.233     &  -2.3 &  2.1 & -5.5 \\
WHO Status        & $<\!0.001$ & 0.447     & $<\!0.001$ &  37.2 & -4.3 & 69.2 \\
Sex               & 0.835     & 0.627     & 0.851     &  -2.0 & -1.8 & -3.7 \\
Smoker            & 0.684     & 0.214     & 0.037     &  -4.5 & -1.5 &  1.4 \\
Cell Type         & 0.019     & 0.835     & 0.002     &   4.0 & -5.1 &  8.6 \\
Metastases        & $<\!0.001$ & 0.056     & $<\!0.001$ &  29.6 &  1.8 & 50.8 \\
Sodium            & 0.009     & 0.154     & 0.002     &   5.4 & -0.3 &  9.4 \\
Albumen           & $<\!0.001$ & 0.251     & $<\!0.001$ &  13.4 & -1.2 & 18.5 \\
&&&&&&\\[-0.3cm]
\hline
&&&&&&\\[-0.4cm]
\multicolumn{7}{l}{\small {\it Note:} $\Delta_{\text{AIC}} = \text{AIC}_{\text{reduced}} - \text{AIC}_{\text{full}}$.}
\end{tabular}
\end{table}

{The results are shown in Table \ref{tableLRandAIC}. Both the p-values arising from the likelihood ratio tests and the AIC differences agree, i.e., those with smaller p-values have positive $\Delta_{\text{AIC}}$ values and vice versa.  It is interesting to note that although age is significant in the shape component, the overall (joint) effect of age is non-significant. Conversely, smoking status is not significant in either component, when judged individually, but the joint effect \emph{is} significant. These findings are a consequence of the correlation that exists between scale and shape coefficients (discussed in Section \ref{hyptestsec} and Appendix A) and, furthermore, the need for considering scale, shape and joint effects is clear.} {The joint $\Delta_{\text{AIC}}$ values give us a sense of the relative importance of each covariate and, in particular, WHO status, treatment and the presence of metastases appear to be the most important factors. Cancer cell type, sodium level and albumen level are also significant, but to a lesser degree. Finally, smoking status has a weak effect whereas age and sex have little impact on survival.}

\begin{table}[!htbp]
\caption{Selection Procedure Results\label{tablevarselect}}
\centering
\begin{tabular}{l@{\hspace{1cm}}c@{\hspace{1cm}}c@{\hspace{1cm}}c@{\hspace{1cm}}c}
\hline
&&&&\\[-0.4cm]
& \multicolumn{2}{c}{\hspace{-0.7cm}\underline{AIC Selection}} &  \multicolumn{2}{c}{\underline{BIC Selection}} \\
 &  PH  & MPR & PH & MPR  \\
\hline
&&&&\\[-0.3cm]
Treatment          & $\beta$  & $\beta,\alpha$  & $\beta$  & $\beta,\alpha$  \\
Age Group          & ---      & ---             & ---      & ---             \\
WHO Status         & $\beta$  & $\beta$         & $\beta$  & $\beta$         \\
Sex                & ---      & ---             & ---      & ---             \\
Smoker             & $\beta$  & $\alpha$        & ---      & ---             \\
Cell Type          & $\beta$  & $\beta$         & $\beta$  & $\beta$         \\
Metastases         & $\beta$  & $\beta,\alpha$  & $\beta$  & $\beta$  \\
Sodium             & $\beta$  & $\beta$         & $\beta$  & $\beta$  \\
Albumen            & $\beta$  & $\beta,\alpha$  & $\beta$  & $\beta$  \\
&&&&\\[-0.3cm]
\hline
&&&&\\[-0.3cm]
AIC                & $3723.1$ & $3679.7$        & ---      & ---     \\
BIC                & ---      & ---             & $3816.7$ & $3802.4$ \\
&&&&\\[-0.4cm]
\hline
&&&&\\[-0.4cm]
\multicolumn{5}{l}{\small {\it Note:} $\beta$ = ``selected in scale'' and $\alpha$ = ``selected in shape'' }
\end{tabular}
\end{table}

Our stagewise selection procedure (based on both AIC and BIC with scale, shape and simultaneous steps) was applied to the full MPR model  and, for comparison, the full PH model (which, of course, only had scale steps). The results are summarized in Table \ref{tablevarselect}. Both the PH and MPR models are in agreement on the significant covariates where, in particular, age and sex are not selected; smoking status is also not selected using BIC. Although the PH and MPR models are in agreement on which covariate effects are significant, the \emph{nature} of the effects will differ under the two models. Specifically, any covariate with an $\alpha$ effect in the MPR model is judged to have a non-PH effect. Thus, using AIC, treatment, smoking status, metastases and albumen level are judged to be non-PH whereas, using BIC, only treatment is judged to be non-PH. Thus, there is strong evidence that the effect of treatment is non-PH which is also supported by the large $\Delta_{\text{AIC}}$ value for its shape effect in Table \ref{tableLRandAIC}. 
The \emph{adjusted} treatment hazard ratios from this multi-factor MPR model (not shown) are quite similar to the unadjusted hazard ratios in Fig.~\ref{hrplot} although the effects of surgery and radiotherapy are reduced (i.e., the hazard ratios are closer to one).

\section{Discussion \label{discuss}}
Multi-parameter regression produces flexible parametric models which can, among other things, relax the proportional hazards assumption by allowing time-dependent hazard ratios. Furthermore, this is achieved without the need for specialized estimation procedures. Naturally, the process of variable selection is more involved in this multi-component setting due to correlation between estimated regression coefficients -- an aspect which does not appear to have been considered in detail by other authors. However, our proposed selection procedure generalizes standard approaches (see Appendix B). Although we have focussed only on the two-parameter Weibull model in this paper, the use of alternative models is advisable in practice. In particular, more general parametric models can facilitate more general functional forms for the hazard ratio. Accordingly, the \texttt{mpr} package \citep{burke:2016} includes a variety of distributions. Finally, we believe that it is important for practitioners to diversify beyond standard Cox analyses and we suggest that multi-parameter regression models provide a flexible alternative.

\section*{Acknowledgements}

We greatly appreciate the insightful comments made by the referees which allowed us to improve on an earlier draft of this paper. The first author would like to thank the Irish Research Council (www.research.ie) for supporting this work.

\newpage

\bibliographystyle{apalike}
\bibliography{refs}

\appendix

\renewcommand\thefigure{A.\arabic{figure}}
\setcounter{figure}{0}

\section*{\normalsize Appendix A: Correlation between estimated regression coefficients}

In this simulation study we investigate the correlation structure for estimated regression coefficients. The data were simulated from the Weibull MPR model with
\begin{align*}
\log(\lambda) &= -3.0 - 0.2~x_1 - 2.3~x_2 & \log(\gamma) &= -0.2 + 0.1~x_1 + 0.5~x_2
\end{align*}
where $x_1$ and $x_2$ are independent binary covariates. We fix regression coefficients and sample size ($n = 1000$) here as the correlation structure is similar for other values. However, we  varied the proportion censored at three different levels, 20\%, 50\% and 80\%, and each of these simulation scenarios was repeated 500 times. At each repetition the Weibull MPR model was fitted to the simulated data giving a $500 \times 6$ matrix of estimated coefficients for each of the three simulation scenarios.

Figure~\ref{cormatplot} below shows scatter matrices of the estimated regression coefficients for both models in each of the three scenarios. We have not included the intercepts in these scatter matrices as we are primarily interested in coefficients of covariates. We can see that the ``same-covariate-pairs'', $(\hat{\beta}_1, \hat{\alpha}_1)$ and $(\hat{\beta}_2, \hat{\alpha}_2)$, i.e., pairs corresponding to the same covariate, are highly correlated supporting the discussion in Section \ref{hyptestsec}. All other coefficients are relatively uncorrelated, owing to the fact that $x_1$ and $x_2$ are independent in this simulation.

\begin{figure}[!htb]
\centering
\begin{tabular}{c@{\hspace{0.1cm}}c@{\hspace{0.1cm}}c}
{\small 20\% Censored} & {\small 50\% Censored} & {\small 80\% Censored} \\[-0.1cm]
\includegraphics[width=0.3\textwidth, trim = 1cm 1cm 1cm 1cm, clip]{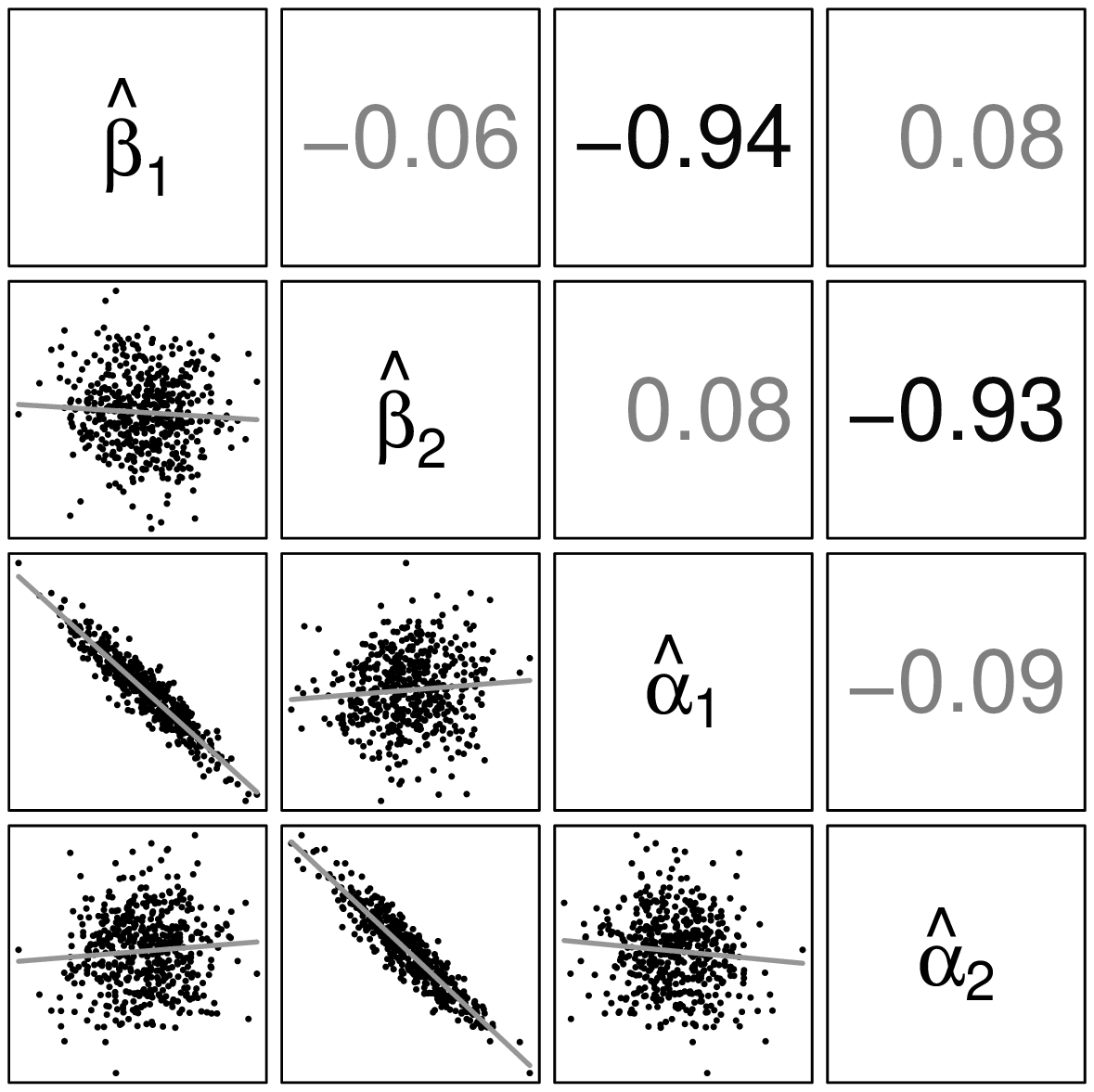} &
\includegraphics[width=0.3\textwidth, trim = 1cm 1cm 1cm 1cm, clip]{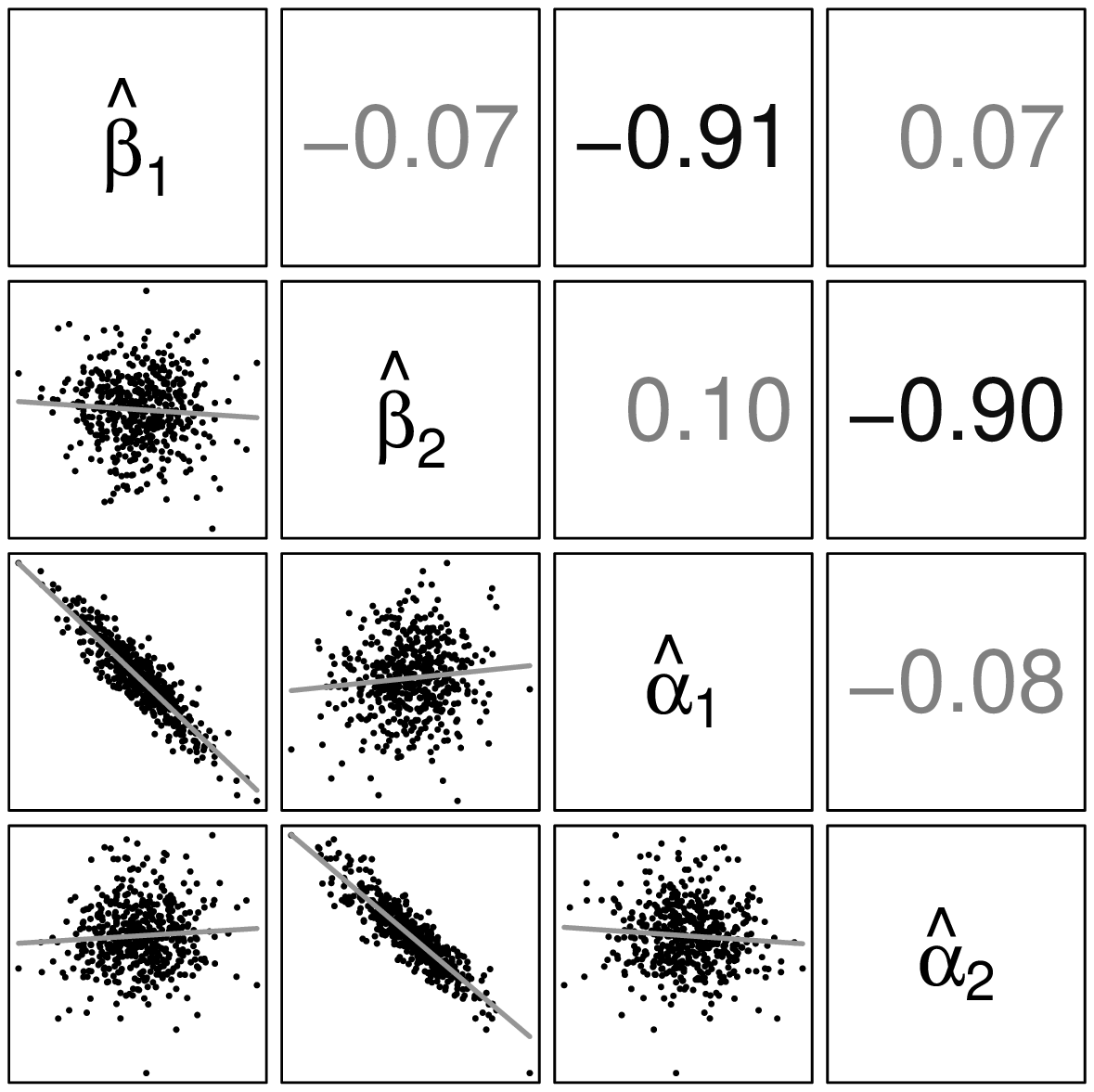} &
\includegraphics[width=0.3\textwidth, trim = 1cm 1cm 1cm 1cm, clip]{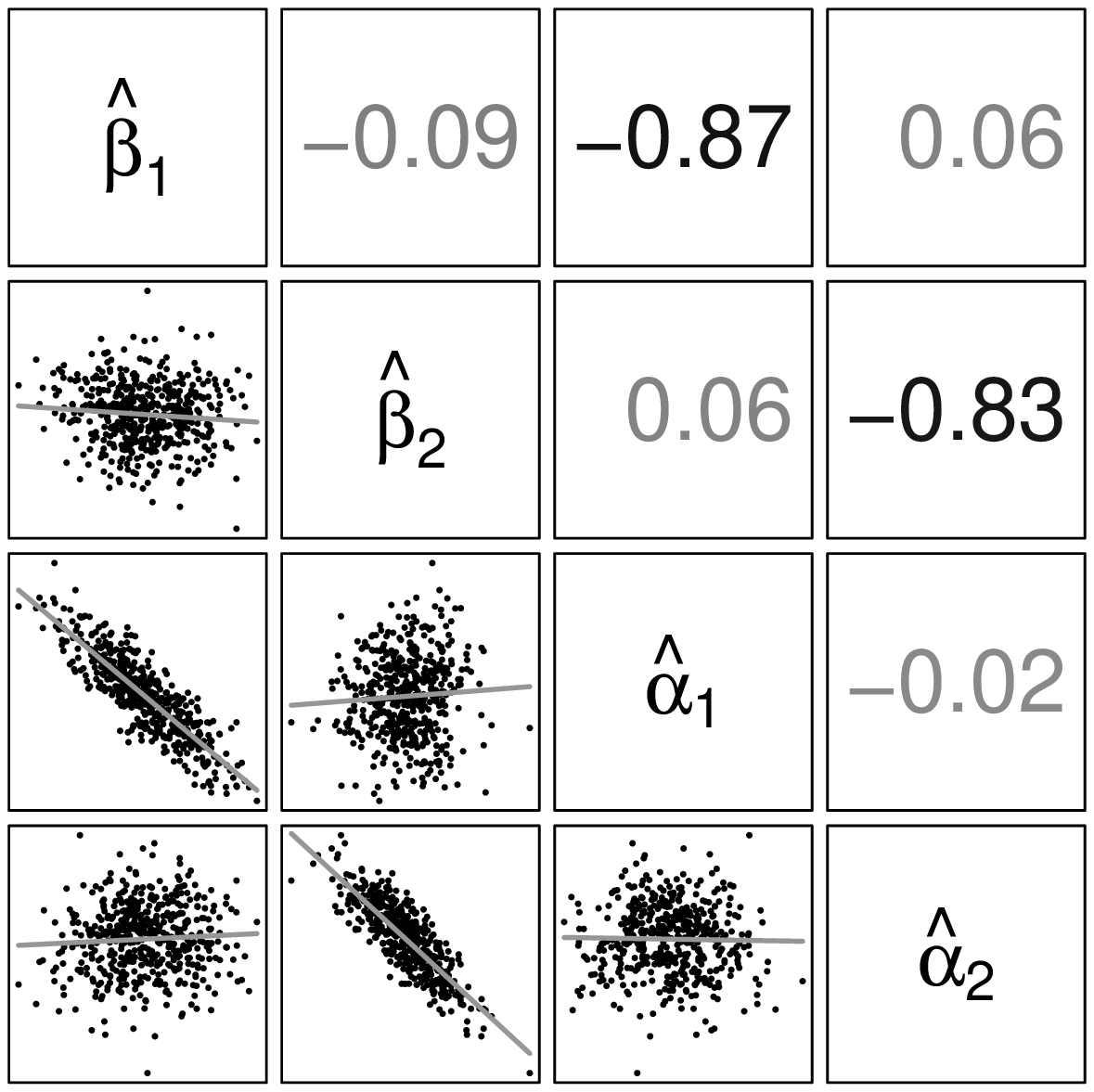}\\[0.6cm]
\end{tabular}
\caption{Correlation matrices based on simulation. The lower triangle shows scatter plots for the estimated regression coefficients with least squares line overlayed. Each scatter plot comprises 500 points arising from 500 simulation repetitions. The upper triangle shows the corresponding correlation value where the larger the correlation, the darker the font.\label{cormatplot}}
\end{figure}

\renewcommand\thefigure{B.\arabic{figure}}
\setcounter{figure}{0}

\section*{\normalsize Appendix B: Multi-parameter regression selection algorithm}
Using the notation of Section \ref{hyptestsec}, we define $M(x,z)$ to be a model with scale and shape covariates $x = (1, x_1, \ldots, x_p)^\T$ and $z = (1, z_1, \ldots, z_q)^\T$ respectively. Figure~\ref{mprfwd} below describes a forward selection procedure starting from the null model $M(x=\emptyset,z=\emptyset)$. At each iteration of the \texttt{while} loop {the algorithm searches for} a model which is better than the current model; ``better'' {in this context} may be based on likelihood ratio tests, Akaike information criterion, bayesian information criterion, etc. {Of course, at each such iteration}, there are candidate covariates to be considered for inclusion in the scale $\{c\,|\,c\not\in x\}$, the shape $\{c\,|\,c\not\in z\}$ and for simultaneous inclusion $\{c\,|\,c\not\in x, z\}$. Hence, {this main loop is composed of} three \texttt{for} loops {where all of these} new models {are fitted} and compared to the current model. If, at the end of the \texttt{while} loop, a better model {is found}, then this becomes the current model and {another iteration of the loop commences}. Otherwise, the loop ends {as the best model has been found.} A more general stagewise procedure has been implemented in the \texttt{mpr} package (the \texttt{stepmpr} function) which includes backward steps in addition to forward steps within the main \texttt{while} loop, i.e., there are three additional \texttt{for} loops carrying out backward steps in the scale, shape and simultaneously in the scale and shape.

\begin{figure}[!htb]
\centering
\includegraphics[width=0.5\textwidth, trim = 4.5cm 7.5cm 9cm 6cm, clip]{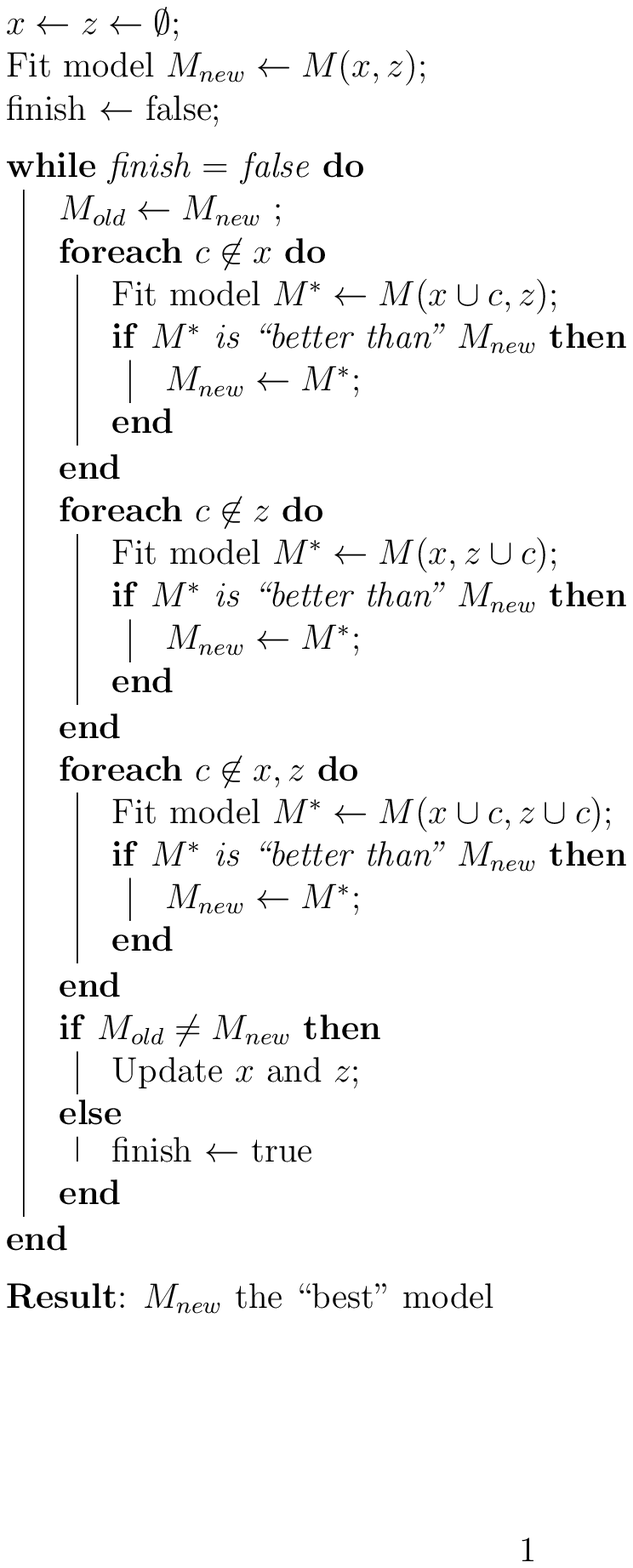}
\caption{Algorithm for multi-parameter regression forward selection. \label{mprfwd}}
\end{figure}

We note that our selection procedure is more general than that which appears in the \texttt{gamlss} package.
Within \texttt{gamlss} there are two strategies: ``A'' and ``B''. Strategy A considers only \emph{one} parameter at a time as follows: first apply forward selection to the scale, then (given the scale model) apply forward selection to the shape and, finally, return to the scale to apply backward selection. Strategy ``B'' in \texttt{gamlss} considers \emph{simultaneous} steps only: apply forward and backward selection simultaneously to both the scale and shape. Our procedure generalizes both of these strategies as, at all points in our selection process, forward and backward steps are applied to both the scale and shape parameters individually and simultaneously.

\end{document}